\documentclass[%
 aip,
 amsmath,amssymb,
 reprint,%
]{revtex4-1}

\usepackage{graphicx}
\usepackage{dcolumn}
\usepackage{bm}

\usepackage[utf8]{inputenc}
\usepackage[T1]{fontenc}
\usepackage{mathptmx}
\usepackage{etoolbox}

\makeatletter
\def\@email#1#2{%
 \endgroup
 \patchcmd{\titleblock@produce}
  {\frontmatter@RRAPformat}
  {\frontmatter@RRAPformat{\produce@RRAP{*#1\href{mailto:#2}{#2}}}\frontmatter@RRAPformat}
  {}{}
}%
\makeatother
\begin{document}

\preprint{AIP/123-QED}

\title{Infrared imaging of samples in ultra high pressure diamond anvil cells}

\author{Tarun Patel}
\affiliation{Institute for Quantum Computing, Department of Physics and Astronomy, and Department of Chemistry, University of Waterloo, Waterloo, ON N2L 3G1, Canada}
\author{A. Drozdov} 
\author{V.S. Minkov}
\author{M.I. Eremets}
\affiliation{Biogeochemistry Department, Max Planck Institute for Chemistry, PO Box 3060, 55020 Mainz, Germany}
\author{E. J. Nicol}
\affiliation{Department of Physics, University of Guelph, Guelph, ON N1G 2W1 Canada}
\author{J. P. Carbotte}
\author{T. Timusk}
\email{timusk@mcmaster.ca}
\affiliation{Department of Physics and Astronomy, McMaster University, Hamilton, ON L8S 4M1, Canada}
\author{\\A. W. Tsen}
\affiliation{Institute for Quantum Computing, Department of Physics and Astronomy, and Department of Chemistry, University of Waterloo, Waterloo, ON N2L 3G1, Canada}

\date{\today}

\begin{abstract}
We describe an experimental platform that generates infrared images of micrometer-sized samples in the high pressure region of a diamond anvil cell.  Using a 2.3 $\mu$m laser as a source of radiation, the system  will be particularly useful in identifying hydride superconductors which exhibit an anomalous temperature dependence of reflectivity in the 2.3 $\mu$m region. Our system shows an intensity stability within one percent when the sample temperature is swept from 100 K to 300 K. The spatial stability is of the order of a few micrometers in the same temperature range.
\end{abstract}

\maketitle

\section{\label{sec:level1}Introduction}

Recent advances in the search for new hydride high temperature superconductors have lead to increasingly  higher pressures and smaller samples in a diamond anvil cell (DAC) environment.\cite{drozdov2015conventional,Drozdov2019superconductivity,somayazulu2019evidence,Eremets2019semimetal}
Conventional techniques that identify the superconducting state such as the Meissner effect, tunnelling, and dc resistivity are difficult with small, possibly inhomogeneous samples. 

\begin{figure}[ht!]
	\vspace*{-0.48 cm}
 	\centerline{\includegraphics[width=10.0cm]{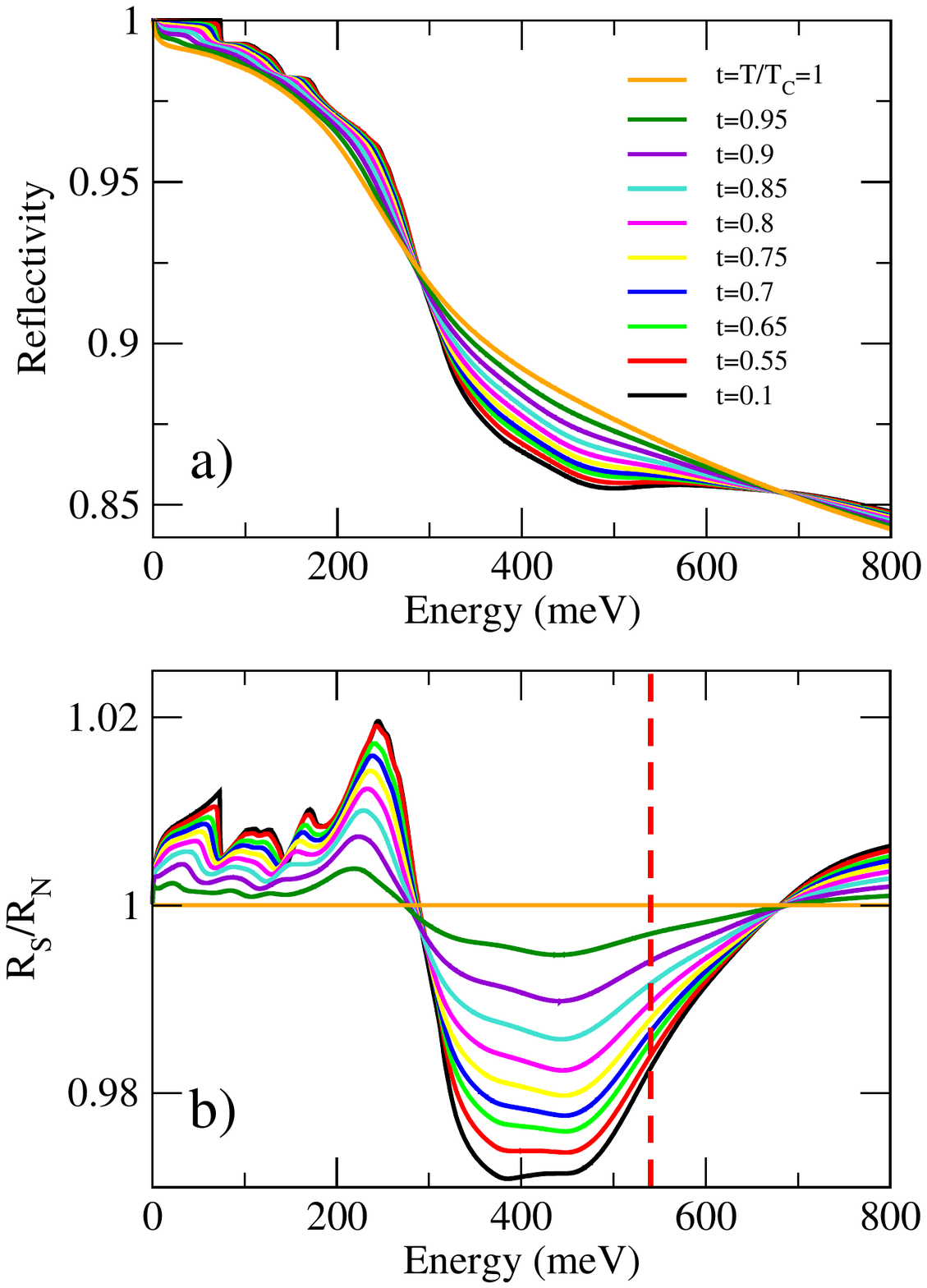}}
	\vspace*{-4.0 cm}%
	\caption{\label{fig:epsart} (a) Theoretical absolute reflectivity of H$_3$S for several temperatures below $T_c$, where $t=T/T_c$ varies from 0.1 to 1. The calculations are in the clean limit with elastic scattering rate $\gamma_r=20$ meV. 			These results are taken from \cite{carbotte2019spectroscopic}. (b) Ratio of the superconducting state reflectivity to that of the normal state at $T_c$ for the temperatures shown in (a). In the range from 300 to 700 meV, the superconducting state is depressed relative to the normal metallic state. This unusual effect, where the sample is more absorbing in the superconducting state, is due to scattering from bosonic excitations at energies between $2\Delta+\hbar\Omega$ and $2\Delta+2\hbar\Omega$, where $\Omega$ is the dominant boson frequency. As shown in our previous work\cite{Capitani2017spectroscopic}, the response in this region is independent of the elastic scattering rate. The vertical red dashed line indicates the 2.3 $\mu$m laser frequency.}
\end{figure}

Infrared optics is a noninvasive probe with micrometer spatial resolution.
Our past work, using a blackbody source, has shown that there is a frequency region in the mid infrared where the superconducting hydrides show an anomalous temperature dependence not seen in ordinary metals. 
This occurs between $2\Delta + \hbar\Omega$ and $2\Delta + 2\hbar\Omega$, where $\Delta$ is the superconducting gap and $\Omega$ is the dominant boson frequency\cite{Capitani2017spectroscopic}.
In this frequency range, the reflectance of a superconductor {\it increases} with increasing temperature, while ordinary metals show a {\it decrease} with rising temperature. 
The overall change in reflectance is of the order of 2 $\%$ between the superconducting and normal states in a typical high pressure hydride.  

Figure 1 shows the calculated reflectance of H$_3$S at different temperatures.
These calculations were done using the standard numerical tools of the BCS-Eliashberg theory of superconductivity with an input electron-phonon spectral density taken from the work of Errea {\it et al.}\cite{Errea:2015}. 
Further details of these and similar calculations for reflectance properties of pressurized hydrides can be found in Refs.~\cite{Capitani2017spectroscopic,carbotte:2018,carbotte2019spectroscopic,Elatresh:2020}.
The upper frame shows the absolute reflectance for several temperatures ranging from $T=0.1T_c$ to $T_c$, where $T_c$ is the superconducting critical temperature.
The lower frame provides the ratio of the reflectance in the superconducting state to that at $T_c$\cite{carbotte2019spectroscopic}.
The red dashed line denotes the frequency of the 2.3 $\mu$m laser.
It is within the range of the anomalous temperature dependence of reflectance, which extends  from 300 meV to 700 meV. 
The overall change in reflectance between the normal state and the fully superconducting state at this frequency is only 2 $\%$. 
At a somewhat lower frequency, the effect is stronger, but practical considerations such as the onset of atmospheric absorption and the availability of lasers favour the 2.3 $\mu$m wavelength.
It is also likely that future, higher T$_c$ materials, will have larger values of $\Omega$, the dominant boson frequency.
Still, to reliably identify superconductivity, the stability of the system has to be better than 0.2 $\%$.
In what follows we describe a system that, with further development, will have the required stability. 

\section{Experimental setup}

Figure 2 shows the overall experimental layout. The diamond anvil cell is mounted on a platform that is thermally coupled to a closed cycle He4 cryostat from Montana Instruments. 
As specified by the manufacturer, the temperature of the platform can be varied from room temperature down to 3.2 K with a position stability of $<$5 nm. 
In our system, the DAC is thermally coupled to the platform with its optical axis vertical. 
A CaF$_2$ window separates the vacuum of the sample space from the ambient environment. 

Our light source is a 4.3 mW, 2.3 $\mu$m diode laser from Eblana Photonics.
It has a built-in thermoelectric cooler, and is driven by a combined dc and ac source.
The ac modulation is at a frequency of 17.8 kHz and is detected with a lock-in amplifier. 
The light from the diode is collimated using an aspheric lens and coupled into a single mode fibre (ZrF4 Patch Cable -Thorlabs). 
The light emitted from the fibre is launched using a silver parabolic mirror. 
Two scanning mirrors in the laser path allow the focal point to be moved in the plane of the culet (the flat high pressure area at the tip of the diamond). 
The beam is focused on the diamond culet with a reflective objective lens, which can be moved along the optic axis under computer control. 
The reflected beam is sent to a battery powered InGaAs infrared detector through a beamsplitter. 
The detector signal is amplified with a preamplifier and sent to the lock-in detector.  
Without any incident laser light on the detector, the lock-in measures 0.4 nW of signal (integration time of 3 ms), indicating a noise level of 0.4 nW. 
The reflected signal from the sample is usually  10 $\mu$W, yielding a noise level of 0.004 $\%$.

\begin{figure*}[ht!]
\centering
\includegraphics[width=13.0cm]{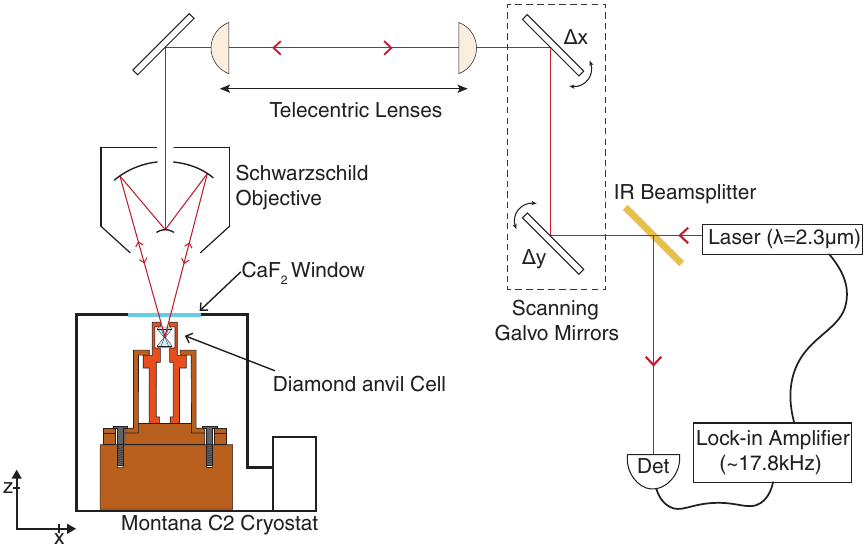}
\caption{Overall layout of the experimental setup. The DAC is mounted inside a Montana Instruments cryostat. The z axis is the direction of the laser beam and the xy plane is the plane of the culet that contains the sample. Various optical elements in the laser path enable xyz positioning of the laser spot on the sample. A reflecting objective is used to focus the laser on the sample. It can be translated along the optic axis (z direction) to keep the image in focus as the sample temperature is varied. Two additional mirrors allow scanning in the xy image plane. The light source is a 2.3 $\mu$m diode laser with a built-in thermoelectric cooler modulated at 17.7 kHz. A battery-powered InGaAs detector is used to detect the reflected signal which is sent through a preamplifier to the lock-in amplifier.}  
\end{figure*}

\begin{figure}[ht!]
\centering
\includegraphics[width= 8cm]{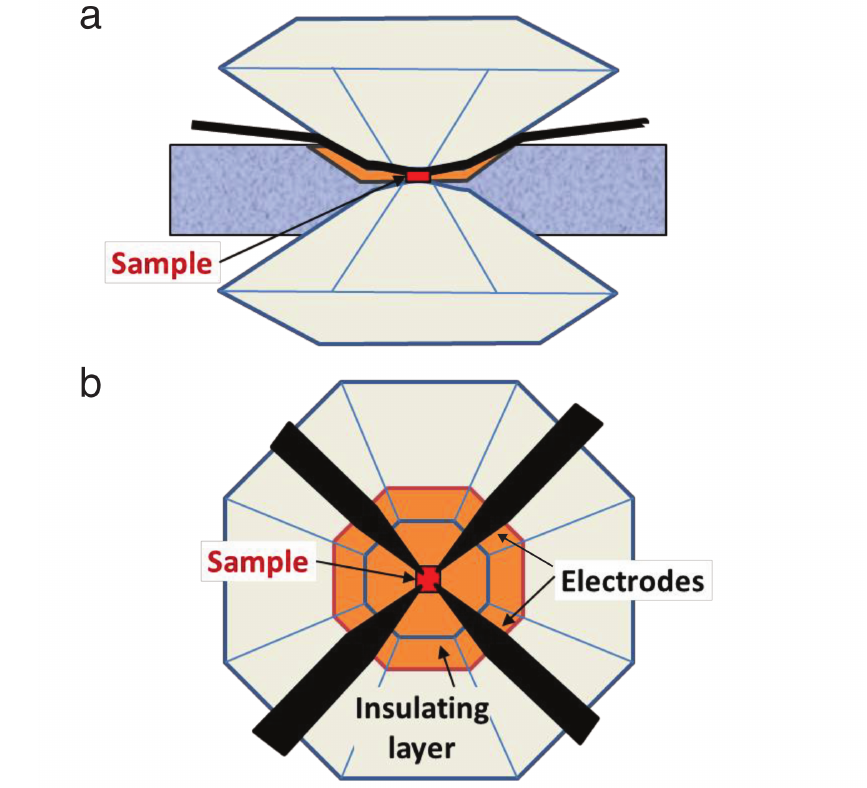}
\caption{
Diamond Anvil Cell
(a)-(b): Schematic view of the used diamond anvil cell (DAC) showing the diamond anvils, the insulating NaCl gasket, the sample and the four electrodes for the resistivity measurements. Note that the electrodes are at an angle to the optic axis so most of the reflected light from them will not reach the detector. Although the platinum electrodes have a higher reflectivity than the sample, in the normal state they appear darker for this reason.  An ideal metallic reference surface would have to be in the same plane as the sample.} 
\label{DAC}
\end{figure}

Figure 3 is a closeup of the diamond anvil cell. 
Panel a) is a side view of the cell.  
The central area is the culet containing the sample, shown in red. 
The sample-diamond surface is flat and normal to the optic axis.
The orange area contains the insulating layer that  insulates the electrodes from the stainless steel gasket, shown in blue. 
The electrodes and the insulating layer are at an angle to the optic axis;  an incident beam on these surfaces is deflected and most of it will not reach the detector. 
For this reason, the electrodes, which are made of high reflectivity metal, appear darker than the sample and cannot be used as a reference surface.  
To get the absolute reflectivity of the sample, a surface with known reflectivity would have to be evaporated on the diamond culet next to the sample. 
However to identify superconductivity it is sufficient to plot the reflected signal as a function of temperature and normalize it to the signal at T$_c$.
T$_c$ can be identified by a kink in the curve where the anomalous temperature changes sign and becomes that of a normal metal\cite{Capitani2017spectroscopic,carbotte2019spectroscopic}.

\section{Results and discussion}

\begin{figure}[ht!]
\centering
\includegraphics[width=8cm]{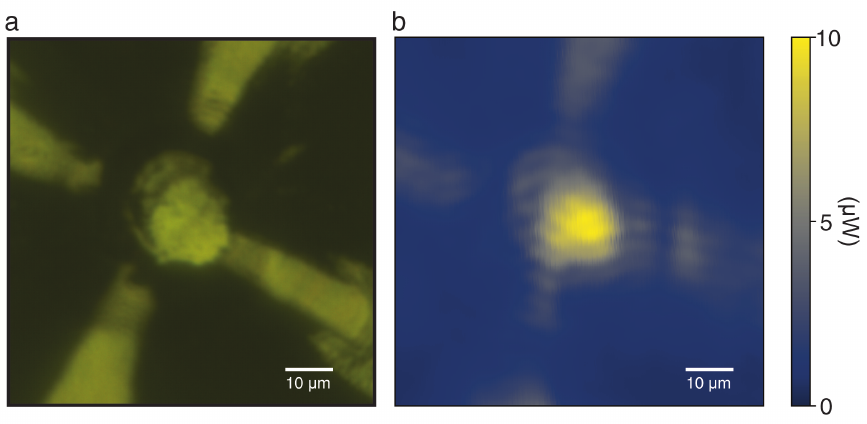}
\caption{ a) Conventional microscope image of the sample region taken with visible light at 300 K. The bright circular spot at the centre is the conducting sample and the four bright arms are the electrodes used to measure the dc resistivity of the sample. b) Scanning-laser image of the same sample at 100 K taken with the 2.3 $\mu$m laser. All the features of the visible image can be recognized in the infrared image. From the loss of sharpness in the infrared image, we estimate a spatial resolution in the xy plane of about 2 $\mu$m. One striking difference in the images is  how much brighter the infrared image is compared to the electrodes.}
\end{figure}

Our goal is to produce high resolution images of the sample region of the DAC that are stable to within a fraction of a percent in amplitude as the temperature of the sample is scanned from deep in the superconducting state at 100 K to the normal state and above as high as 300 K. 
In what follows, we will show that the present system is capable of achieving that goal.

Figure 4 shows two images of the sample area.
On the left, we show an image taken with a conventional visible light microscope at 300 K. 
We see five bright areas: a central spot (the H$_3$S sample) and the four electrodes used to measure the resistivity of the sample in van der Pauw geometry. 
As figure 1 shows, the sample is expected to have a reflectance well below 0.85 in the visible region. 
This poor metal behaviour is due to the  strong electron-phonon interaction of the hydrides.
The platinum electrodes should have a reflectance of the order of 99$\%$ in this frequency region, but because the electrodes are at a finite angle to the xy plane of the sample, most of the reflected signal does not reach the detector and therefore, the electrodes appear darker than the sample. 

The image on the right is taken of the same sample with our infrared imaging system at 100 K using the 2.3 $\mu$m laser as a source.
While all the features of the visible image are present, the image in the infrared is not quite as sharp as in the visible.
What is striking is that the image of the sample is much brighter than the electrodes.

\begin{figure*}[ht!]
\centering
\includegraphics[width=15cm]{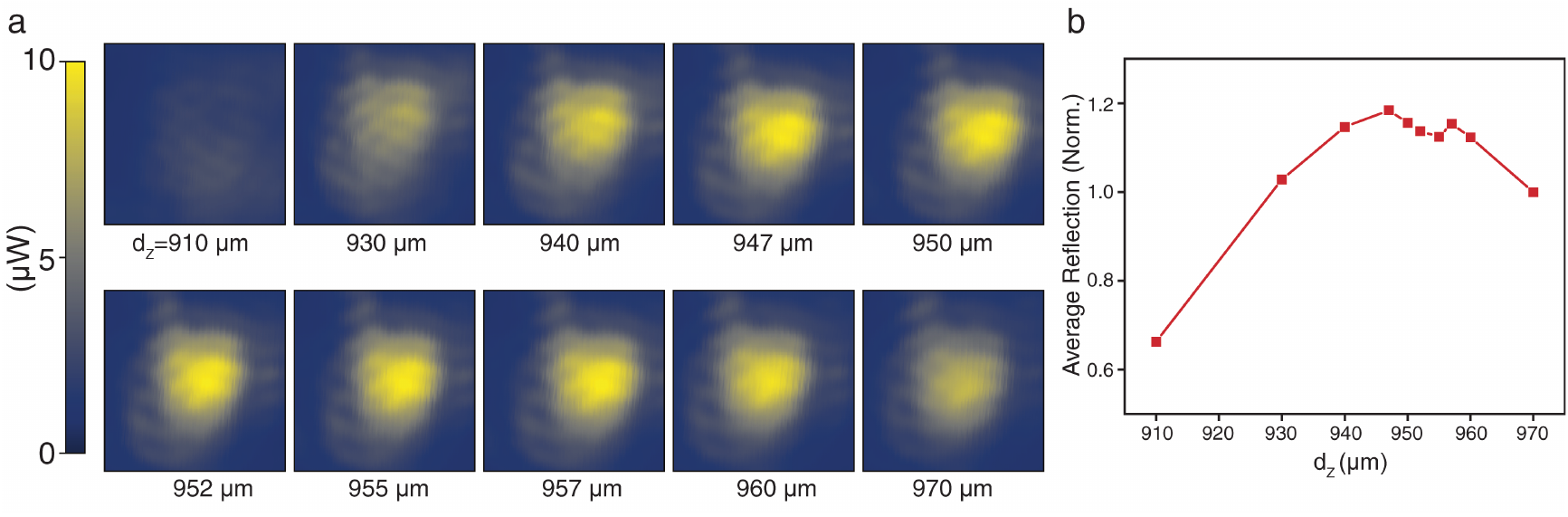}
\caption{ Focusing the laser spot on the sample. As the temperature is varied, the sample moves slightly in the direction of the optic axis. To compensate for this variation, the lens has to be refocussed at each temperature. Panel a) shows the central image at different positions of the objective lens. Each window is $\sim$ 30 $\mu$m x 30 $\mu$m. Note that the image does not shift by more than a few micrometers in the course of several hours used to acquire the images. Panel b) shows the average intensity of the reflected beam vs. the z-position of the lens (see text for details), normalized to the last z-position. The maximum at 945 $\mu$m corresponds to actual focal point.  Such a scan is performed at each temperature to keep the laser focused on the sample.
}
\end{figure*}

In what follows, we discuss the technique used to align the laser and focus the spot on the sample. 
Since the IR laser is invisible to the eye, the optical setup is initially aligned using a $\lambda$ = 650 nm red laser.
The laser spot is focused on the top diamond surface, which is the only available flat surface other than the sample plane for a normal reflection geometry.
The image of the top diamond surface, is then centered in the field of view, and then, the $\lambda$ = 2.3 $\mu$m laser is substituted for the red laser. 
The objective is then lowered by $\sim$ 930 $\mu$m using a motorized z-stage and a reflection image of  $\sim$100 $\mu$m x 100 $\mu$m (much larger than the sample size) is collected.
The objective is then moved in increments of 2-10 $\mu$m with a reflection image recorded after each step.
The sample comes into focus around $\sim$ 945 $\mu$m below the top diamond surface, as seen in figure 5a.
In order to determine the optimal focal plane, we take a $\sim$ 30 $\mu$m x 30 $\mu$m, window centred around the sample region and plot the averaged signal in this region with respect to the displacement of the objective lens (figure 5b). 
Typically, a dome like curve is obtained and the top of the dome is considered to be the optimal focal plane.
The height of the cryostat stage changes with temperature, and early measurements indicate that the top diamond surface moves down $\sim$ 160 $\mu$m as the sample is cooled from 260K to 50K.
Thus, this process needs to be repeated for each temperature point recorded. 

Once the optimal focal plane is determined, the next challenge is to normalize for shifts in the xy plane and the overall intensity of the reflection signal over a temperature sweep. 
Note that figure 5a shows the sample image does not shift in the xy plane by more than a few micrometers in the course of the several hours needed to collect the images, provided the temperature remains constant.
However, as the temperature is changed, the shift will be larger. 
One way to solve this problem is to introduce a metal with a known reflection behavior in the sample plane such as Pt.
The signal from the metal would solve the problem of re-centering and intensity normalization simultaneously. 
An alternate method would be to use the sample itself as a reference point and simply take the maximum of the sample signal as a measure of the reflectance. 
The issue though with such a process, is that the sample would be used for self-calibration, which could obscure important information. 

\section{Conclusions}
In summary, we have developed a system that can measure the reflected intensity of micrometer size features in a diamond anvil cell as a function of temperature with a signal accuracy of four orders of magnitude.
With further improvements it can be used to identify superconductivity in hydrides using lasers tuned to the frequency range where the anomalous temperature dependence is seen. 
While we are using a fixed frequency laser in our current experiments, with a tunable laser a variety of transition temperatures can be identified with our method in the future.
It should be pointed out, however, that the method cannot be used with conventional superconductors. 
While they are expected the show the same anomalous temperature dependence of the reflectivity, the spectral region where this occurs is in the far infrared and the effect is expected to be much smaller.
Correspondingly, the spatial resolution will be of the order of hundreds of micrometers and the samples would have to be of millimeter in size.

\begin{acknowledgments}
AWT acknowledges support from the US Army Research Office (W911NF-21-2-0136), Ontario Early Researcher Award (ER17-13-199), and the Gouvernement du Canada | Natural Science and Engineering Research Council of Canada (RGPIN-2017-03815)(NSERC). This research was undertaken thanks in part to funding from the Canada First Research Excellence Fund.  EJN was supported by NSERC (RGPIN-2017-03931) and TT was supported by NSERC (RGPIN-2018-06865) \\

\end{acknowledgments}

\section*{Data Availability Statement}
The data that support the findings of this study are available within the article.

\nocite{}
\bibliography{jap_1_bib}

\end{document}